\input{aipcheck}

\documentclass[
  ,final            
  ]
  {aipproc}

\layoutstyle{8x11double}

\usepackage{amssymb,epsfig,amsmath}
\usepackage{color}

\begin{document}

\title{$511$ keV $\gamma$-ray emission from the galactic bulge by MeV millicharged dark matter
\footnote{A talk given by J. Park at SUSY'08 (Seoul, Korea, June
16-21, 2008).}}


\author{Ji-Haeng Huh, Jihn E. Kim, Jong-Chul Park, and Seong Chan Park}
{address={Department of Physics and Astronomy, Seoul National
University, Seoul 151-747, Korea} }



\begin{abstract}
We propose a possible explanation for the recently observed
anomalous 511 keV line with a new ``millicharged'' fermion. This
new fermion is light [${\cal O}({\rm MeV})$]. Nevertheless, it has
never been observed by any collider experiments by virtue of its
tiny electromagnetic charge $\varepsilon e$. In particular, we
constrain parameters of this millicharged particle if the $511$
keV cosmic $\gamma$-ray emission from the galactic bulge is due to
positron production from this new particle.
\end{abstract}

\maketitle


\section{Introduction}
After the first detection of $\sim500$ keV $\gamma$-rays from the
galactic center (GC)~\cite{FirstObserv}, various observations have
been performed. The recent SPI/INTEGRAL observation shows the very
sharp peak from the galactic bulge, which is very well consistent
with the electron mass $m_e\simeq511$
keV~\cite{Knodlseder:2003sv,spectral}. The morphology of the
emission region is consistent with a 2-dimensional gaussian of a
full width at half maximum of $6^\circ$ with a 2$\sigma$
uncertainty range, $4^\circ-9^\circ$. The most probable
explanation of this line emission is that it comes from the
positronium decay. Therefore, a sufficient number of positrons are
needed. Some obvious candidate origins for positrons are the
astrophysical sources such as massive stars, hypernovae,
cosmic-ray interactions, X-ray binaries, type Ia supernovae.
However, these sources are inappropriate to explain the intensity
of 511 keV $\gamma$-ray flux and the shape of emission region.
Thus, particle physics origins with new particles are currently
more preferred. One of them is low mass ($\sim$ MeV) particle dark
matter (DM) annihilation~\cite{Boehm:2003bt}. Along this idea, we
propose a new DM candidate with $\cal{O}$(MeV) mass and very small
electromagnetic charge,\footnote{Only neutral particles are
typically considered as DM candidates. However, charged particles
also could be a good candidate if their electric charge is
sufficiently tiny.} which can explain the galactic 511 keV
$\gamma$-rays.

If there exists another massless $U(1)$ gauge boson, ``{\it
exphoton},'' beyond the standard model (SM), most probably a
kinetic mixing is generated via loop effects between the photon
and exphoton. After a proper diagonalization procedure of kinetic
energy terms, hidden sector particles can be electromagnetically
millicharged~\cite{Okun:1982xi}.\footnote{The term
``millicharged'' does not mean really a milli-size electromagnetic
charge but just a small charge.} Laboratory and cosmological
bounds of millicharged particles are well summarized in
Ref.~\cite{Davidson:2000hf}, but some constraints such as the 511
keV photon flux and the Debye screening are not included.
Therefore, we investigate the possibility of the $\mathcal{O}$
(MeV) millicharged particles toward interpreting the 511 keV line
emission and some related constraints~\cite{JCPark}.

\section{Kinetic mixing and millicharged particle}
For $U(1)_{\rm em}$ and $U(1)_{\rm ex}$ symmetries, the kinetic
mixing of $U(1)_{\rm em}$ and $U(1)_{\rm ex}$ gauge bosons is
parameterized as
\begin{eqnarray}
{\cal L} = -\frac{1}{4}\hat{F}_{\mu\nu}\hat{F}^{\mu\nu}
-\frac{1}{4}\hat{X}_{\mu\nu}\hat{X}^{\mu\nu}
-\frac{\xi}{2}\hat{F}_{\mu\nu}\hat{X}^{\mu\nu}\;,
\label{U(1)emU(1)ex}
\end{eqnarray}
where $\hat{A}_\mu$ and $\hat{X}_{\mu}$ are $U(1)_{\rm em}$ and
$U(1)_{\rm ex}$ gauge bosons, and their field strength tensors are
$\hat{F}_{\mu\nu}$ and $\hat{X}_{\mu\nu}$ respectively. Although
the kinetic mixing parameter $\xi$ is expected to be generated by
an ultraviolet theory~\cite{Okun:1982xi}, $\xi$ can be treated as
a arbitrary parameter in a low energy effective theory. A proper
transformation of the gauge fields,
\begin{eqnarray}
\left(
  \begin{array}{c}
    A_\mu \\
    X_\mu \\
  \end{array}
\right) = \left(
            \begin{array}{cc}
              \sqrt{1-\xi^2} & 0 \\
              \xi & 1 \\
            \end{array}
          \right) \left(
                    \begin{array}{c}
                      \hat{A}_\mu \\
                      \hat{X}_\mu \\
                    \end{array}
                  \right)\;,
\end{eqnarray}
leads to
\begin{equation}
{\cal L} =
-\frac{1}{4}F_{\mu\nu}F^{\mu\nu}-\frac{1}{4}X_{\mu\nu}X^{\mu\nu}\;,
\end{equation}
where photon and {\it exphoton} correspond to $A_\mu$ and $X_\mu$
respectively, and $F_{\mu\nu}$ and $X_{\mu\nu}$ are the new field
strengths.

Let us take the following interaction Lagrangian of a SM fermion,
i.e. electron, in the original basis as
\begin{equation}
{\cal L}= \bar{\psi} \left( \hat{e} Q \gamma^\mu \right) \psi
\hat{A}_\mu\;. \label{SM-interaction}
\end{equation}
If there is a hidden sector fermion $\chi$ with a $U(1)_{\rm ex}$
charge $Q_{\chi}$, its interaction with the hidden gauge boson is
given by
\begin{equation}
{\cal L}= \bar{\chi} \left( \hat{e}_{\rm ex} Q_{\chi} \gamma^\mu
\right) \chi \hat{X}_{\mu}\;. \label{hidden-interaction}
\end{equation}
Note that no direct interaction exists between the electron and
the hidden gauge boson $\hat{X}$, and also between the hidden
fermion and the visible sector gauge boson $\hat{A}$.

In the transformed basis, Eq.~(\ref{SM-interaction}) is rewritten
as
\begin{equation}
{\cal L}= \bar{\psi} \left( \frac{\hat{e}}{\sqrt{1-\xi^2}} Q
\gamma^\mu \right) \psi A_\mu\;.
\end{equation}
Even after the change of basis, the SM fermion has a coupling only
to the photon $A$. On the other hand, the coupling $\hat{e}$ is
modified to $\hat{e} / \sqrt{1-\xi^2}$; consequently, the physical
visible sector coupling $e$ is defined as $e \equiv \hat{e} /
\sqrt{1-\xi^2}$. Similarly, we obtain the following for $\chi$,
\begin{equation}
{\cal L} = \bar{\chi} \gamma^\mu \left(\hat{e}_{\rm ex} Q_{\chi}
X_\mu -\hat{e}_{\rm ex}\frac{\xi}{\sqrt{1-\xi^2}} Q_{\chi} A_\mu
\right) \chi\;. \label{hidden-shift}
\end{equation}
In this basis, the hidden fermion $\chi$ can couple to the photon
$A$ with the coupling $-\hat{e}_{\rm ex} \xi / \sqrt{1-\xi^2}$.
Thus, we set the physical hidden coupling as $e_{\rm ex} \equiv
\hat{e}_{\rm ex}$ and define the coupling of $\chi$ to the photon
$A$ as $\varepsilon e \equiv -{e}_{\rm ex} \xi / \sqrt{1-\xi^2}$
by introducing the millicharge parameter $\varepsilon$. Note that
$e\ne e_{\rm ex}$ in general. In principle, one can calculate the
ratio $e_{\rm ex}/e$ from a fundamental theory. However, we simply
take the ratio as a free parameter.

\section{Constraints on light millicharged dark matter}
\begin{figure}[!t]
\includegraphics[width=7.8cm]{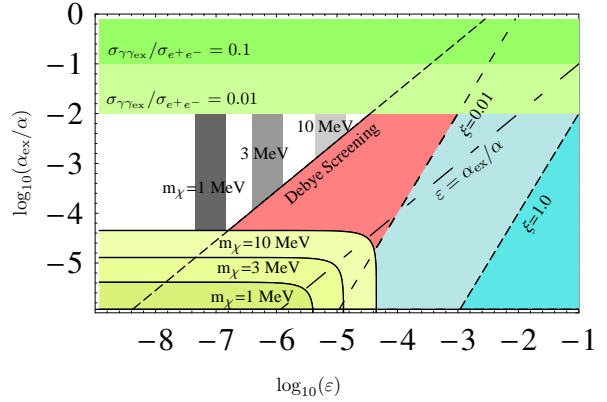}
\caption{The plot for $\alpha_{\rm ex}/\alpha$ versus
$\varepsilon$.} \label{relic}
\end{figure}

Needed cross sections for the cosmological study of $\chi$ are
$\chi\bar{\chi}\rightarrow 2\gamma_{\rm ex}$,
$\chi\bar{\chi}\rightarrow e^-e^+$, $\chi\bar{\chi}\rightarrow
\gamma\gamma_{\rm ex}$, and $\chi\bar{\chi}\rightarrow \gamma
\gamma$. The ratio for these cross sections is given by
\begin{eqnarray}
\sigma_{2\gamma_{\rm ex}}:\sigma_{e+e-}:\sigma_{\gamma \gamma_{\rm
ex}}:\sigma_{2\gamma} \simeq  \alpha_{\rm ex}^2: \varepsilon^2
\alpha^2:\varepsilon^2 \alpha\alpha_{\rm ex}:\varepsilon^4
\alpha^2.
\end{eqnarray}
The first two channels mainly determine the relic density of
$\chi$ since the last two channels are quite suppressed in the
parameter region where $\varepsilon$ and $\alpha_{\rm ex}/\alpha$
are {\it small} as is required by observational data. If
$\alpha_{\rm ex}/\alpha > 0.01(0.1)$, the background diffuse
$\gamma$-ray flux could be larger than $1(10)\%$ of the 511 keV
flux, so the region is already excluded~\cite{Comptel,
DiffuseBack} (see the upper (green) region in Fig.~\ref{relic}).
The second process determines the 511 keV flux as well.

The relic density of a generic relic, $X$, is given by
\begin{equation}
\Omega_X h^2 \approx \frac{1.07 \times 10^9 \, {\rm GeV}^{-1}}
{M_{Pl}}\frac{x_F}{\sqrt{g_*}}\frac{1}{(a+3b/x_F)}\;,
\end{equation}
where $g_*$ is evaluated at the freeze-out temperature $T_F$,
$x_F=m_X/T_F \simeq17.2 + \ln(g/g_*)+\ln(m_X/{\rm
GeV})+\ln\sqrt{x_F}$ for 1 MeV$\lesssim m_X \lesssim$1 GeV, and
the cosmological average of the cross section times velocity is
expressed as $\langle\sigma v\rangle = a+b\langle v^2\rangle+{\cal
O}(\langle v^4\rangle)$~\cite{Bertone:2004pz}. Using
$a=a_{e^-e^+}+a_{2\gamma_{\rm ex}}$, $b=b_{e^-e^+}+b_{2\gamma_{\rm
ex}}$, and an approximated relation $x_F \approx 11.6+\ln(m_X/{\rm
MeV})$ for 1 MeV$\lesssim m_X \lesssim$100 MeV, we can estimate
the relic density of the millicharged particle $\chi$ as
\begin{equation}
\Omega_{\chi} h^2 \approx \frac{1.60 \times 10^{-13} \, (11.6+\ln
\overline{m}) \overline{m}^2}{\left(\frac{\alpha_{\rm ex}}{\alpha}
\right)^2+\varepsilon^2\left(1-\frac{m^2_e}{m^2_{\chi}}
\right)^{1/2}\left(1+\frac{m^2_e}{2m^2_{\chi}}\right)}\;,
\end{equation}
where $\overline{m}\equiv m_{\chi}/{\rm MeV}$ and we put
$g_*\simeq 10.75$ for $1<T_F/{\rm MeV}<100$. Finally, we find a
constraint for $m_{\chi}$, $\varepsilon$, and $\alpha_{\rm
ex}\equiv e^2_{\rm ex}/4\pi$, based on the WMAP three-year
results~\cite{Spergel:2006hy}. In Fig.~\ref{relic}, the lower left
(yellow) corner is excluded by our DM relic density analysis: the
lines correspond to $\Omega_{\chi}h^2=0.11$ and $m_{\chi}=1, 3,$
and $10$ MeV, respectively. Recent analysis such as internal
bremsstrahlung radiation and in-flight annihilation gives strong
mass bound for DM in MeV region: $m \lesssim 3-4$
MeV~\cite{DiffuseBack}. This bound can be reduced by a factor of
two by a possible ionization of the medium~\cite{lowerm}.
Therefore, in this study we focus on the mass range,
$m_\chi\lesssim10$ MeV.

If $m_\chi < m_\mu$, the low velocity annihilations dominantly
produce $e^-e^+$ pairs. Most positrons lose energy through their
interactions with the inter stellar medium and bremsstrahlung
radiation, and go rest. Thus, positron annihilation takes place
via the positronium formation ($\sim 96.7\pm 2.2
\%$)~\cite{spectral} and partly via the direct annihilation into
two $511$ keV $\gamma$-rays. A singlet positronium state decays to
two 511 keV photons ($25\%$), whereas a triplet state decays to
three continuum photons ($75\%$). As a result, the $511$ keV
$\gamma$-ray flux from the galactic bulge can be given as
\begin{equation}
\begin{split}
\Phi_{\gamma,511} \simeq\; &0.275 \times 5.6\;
\overline{J}(\Delta \Omega) \Delta \Omega\\
&\times \bigg(\frac{\sigma v}{{\rm pb}}\bigg)
\bigg(\frac{1\,\rm{MeV}}{m_{\chi}}\bigg)^2\; {\rm cm}^{-2} {\rm
s}^{-1}\;, \label{flux}
\end{split}
\end{equation}
where $\Delta \Omega$ is the observed solid angle toward the GC
and $\overline{J}(\Delta \Omega)$ is defined as the average of
$J(\psi)$ over a spherical region of solid angle $\Delta\Omega$
centered on $\psi=0$~\cite{Bertone:2004pz}. Here, we separate halo
profile depending factors from particle physics depending factors
by introducing the quantity $J(\psi)$ :
\begin{equation} J\left(\psi\right) = \frac{1}
{8.5\, \rm{kpc}} \left(\frac{1}{0.3\,
\mbox{\small{GeV/cm}}^3}\right)^2 \int_{\mbox{\small{l.o.s}}}d s
\rho^2\left(r(s,\psi)\right),
\end{equation}
where $s$ is a coordinate running along the line of sight (l.o.s)
in a direction making an angle $\psi$ from the direction of the
GC.

$\Phi_{\gamma,511} \simeq (1.02 \pm 0.10) \times 10^{-3}$ ph
cm$^{-2}$ s$^{-1}$~\cite{Knodlseder:2003sv,spectral} and $e^+$ is
produced from the process $\chi\bar{\chi}\rightarrow e^-e^+$.
Therefore, considering a solid angle of $0.0086$ sr corresponding
to a $6^{\circ}$ diameter circle, we can find the charge
$\varepsilon$ of the millicharged DM as a function of its mass
$m_{\chi}$ :
\begin{eqnarray}
\varepsilon \simeq 1.0
\times10^{-6}\frac{\overline{m}^2}{\sqrt{\overline{J}}}
\left[1-\frac{m^2_e}{m^2_{\chi}}\right]^{-1/4}
\left[1+\frac{m^2_e}{2m^2_{\chi}}\right]^{-1/2}.
\end{eqnarray}
To estimate the required parameter space, we use the width of the
observed distribution $\overline{J}(0.0086{\rm\ sr}) \sim 50-500$,
approximately corresponding to $\gamma \simeq 0.6-1.2$,
essentially following the approach of
Ref.~\cite{Boehm:2003bt}.\footnote{If the main source of 511 keV
$\gamma$-rays from galactic bulge is the DM annihilation, the
observed distribution of 511 keV emission line would constrain the
shape of the DM halo profile because DM annihilation rate is
proportional to the DM density squared.} In Fig.~\ref{relic}, we
show the allowed range of $\varepsilon$ for typical DM masses
($m_\chi = 1, 3$, and $10$ MeV) as the (grey) vertical bands,
obtained from the 511 keV $\gamma$-ray flux analysis.

If DM is charged, photon obtains effective mass in the DM plasma,
and this mass should be smaller than the experimental limit. As a
result, the Debye screening length in the DM plasma around Earth
$\lambda_D= \sqrt{T_\chi/ \varepsilon^2 e^2 n_\chi}$ is required
to be larger than $1/m_\gamma^{\rm eff}$~\cite{Debye}. Putting
$n_\chi = \rho_\chi/m_\chi \simeq 0.3 {\rm GeV/cm^3} \times
\Omega_\chi/(\Omega_{\rm DM}m_\chi)$ and $\Omega_{\rm DM}\simeq
0.23$, we finally obtain a simple relation $\frac{\alpha_{\rm
ex}}{\alpha} \gtrsim 282 \varepsilon$. The lower right corner
(pink) from the central region is excluded by this constraint.

In the upper part of the line $\varepsilon = \alpha_{\rm
ex}/\alpha$ the process $\chi\bar{\chi}\rightarrow 2\gamma_{\rm
ex}$ and in the lower part the process $\chi\bar{\chi}\rightarrow
e^-e^+$ dominates respectively. Therefore, in the allowed
parameter region ($1 \gg \alpha_{\rm ex}/\alpha > \varepsilon$),
the relic density of DM is essentially determined by
$\chi\bar{\chi}\rightarrow 2\gamma_{\rm ex}$. However, the
observed $511$ keV photon flux is mostly explained by
$\chi\bar{\chi}\rightarrow e^-e^+$. In this respect, the
difficulty of explaining both quantities in
Ref.~\cite{Boehm:2003bt} is easily evaded in our model. As can be
seen from Fig.~\ref{relic}, a significant region is excluded.
However, we note that there still remains an available space.

\section{Conclusion}
We suggested the MeV millicharged dark matter as a possible
solution for the recently observed anomalous 511 keV cosmic
$\gamma$-rays. In this regard, we considered various bounds
including the relic density, the Debye screening, and the diffuse
$\gamma$-ray background. From this study, we conclude that the
millicharged particle hypothesis is not ruled out yet but there
remains only a small parameter space compatible with the 511 kev
$\gamma$-ray flux. Finally, we note that a millicharged particle
with a small mass is preferred as long as its mass is larger than
the electron mass $m_e$ for it to constitute a sizable portion of
the dark matter content of the Universe.




\begin{theacknowledgments}
This work was supported in part by the Korea Research Foundation
Grant funded by the Korean Goverment (MOEHRD)
(KRF-2005-084-C00001).
\end{theacknowledgments}




\bibliography{sample}



\end{document}